\documentclass[prl,twocolumn,showpacs,aps]{revtex4}
\usepackage{graphicx,graphics,color,epsfig}
\usepackage{bm}
\usepackage{amsmath}
\usepackage{amssymb}
\usepackage{epstopdf}

\makeatletter

\newcommand{\Rmnum}[1]{\expandafter\@slowromancap\romannumeral #1@}
\makeatother

\begin{document}

\title{Pairing symmetry in BiS$_{2}-$based superconductors}

\author{Yi Gao}
\affiliation{Department of Physics and Institute of Theoretical Physics,
Nanjing Normal University, Nanjing, Jiangsu, 210023, China}

\begin{abstract}
The possible pairing symmetries for BiS$_{2}-$based superconductors is investigated by using a minimal two-orbital model with onsite and nearest-neighbor intraorbital attractions $V_{0}$ and $V_{1}$, respectively. By using the mean-field approximation and solving the self-consistent equations, the phase diagram of the pairing symmetry is obtained. It is shown that the model allows three possible pairing symmetries, depending on the values of $V_{0}$ and $V_{1}$: the isotopic $s-$wave pairing [$\Delta_{\mathbf{k}}=\Delta_{s}$], the anisotropic $s-$wave pairing [$\Delta_{\mathbf{k}}=\Delta_{s}+\frac{\Delta_{xs}}{2}(\cos k_{x}+\cos k_{y})$] and the $d-$wave pairing [$\Delta_{\mathbf{k}}=\frac{\Delta_{d}}{2}(\cos k_{x}-\cos k_{y})$]. Furthermore the density of states for these pairing symmetries exhibit different behaviors which can be used to distinguish them.
\end{abstract}

\pacs{74.70.-b, 74.20.Rp, 74.25.-q}

\maketitle

\emph{Introduction}.---The recently discovered family of BiS$_{2}-$based superconductors has attracted much attention due to its similarity with the cuprates and iron pnictides. It displays a layered structure where superconductivity is believed to occur within the BiS$_{2}$ plane, similar to the CuO and FeAs planes in the cuprates and iron pnictides, respectively. Superconductivity with $T_{c}=4.5$K was first reported in Bi$_{4}$O$_{4}$S$_{3}$ \cite{1}. Later it was found that $Re$O$_{1-x}$F$_{x}$BiS$_{2}$ ($Re=$La, Nd, Ce and Pr) can also exhibit superconductivity \cite{2,3,4,5} with the highest $T_{c}=10.6$K reported in LaO$_{0.5}$F$_{0.5}$BiS$_{2}$ \cite{2}. These findings suggest that the BiS$_{2}-$based superconductors can also have relatively high transition temperature and it is of great importance to understand the superconducting (SC) pairing mechanism and symmetry in this kind of materials, since studying these may help to unravel the mystery of the pairing mechanism in high-temperature superconductors.

The band structure of this kind of materials has been calculated by first principles calculation \cite{6,7,8}, where the energy bands close to the Fermi level can be reproduced by a simplified two-orbital model \cite{6}. It was shown that around $x\approx0.5$, the Fermi surface topology changes and the good nesting of the Fermi surface in this case may be the cause of the high $T_{c}$. Meanwhile the SC symmetry is predicted to be $s-$wave with a constant gap sign if the electron-phonon coupling is important, whereas a sign-reversing $s-$wave gap can be obtained if the spin fluctuation plays the main role in the Cooper pairing. In addition, other pairing symmetries have also been proposed \cite{9,10,11}.

In this paper, we study the possible pairing symmetries in a minimal two-orbital model with onsite and nearest-neighbor (NN) intraorbital attractive interactions $V_{0}$ and $V_{1}$, respectively. By using the mean-field approximation, the phase diagram of the pairing symmetry is obtained. We found that in this model there exist three possible pairing symmetries, depending on the values of $V_{0}$ and $V_{1}$. The first one is the isotopic $s-$wave pairing [$\Delta_{\mathbf{k}}=\Delta_{s}$]. The second one is the anisotropic $s-$wave pairing [$\Delta_{\mathbf{k}}=\Delta_{s}+\frac{\Delta_{xs}}{2}(\cos k_{x}+\cos k_{y})$] and the last one is the $d-$wave pairing [$\Delta_{\mathbf{k}}=\frac{\Delta_{d}}{2}(\cos k_{x}-\cos k_{y})$]. Furthermore we propose that the density of states (DOS) for these pairing symmetries exhibit different behaviors which can be used to distinguish them.

\emph{Method}.---We begin with the minimal two-orbital model for BiS$_{2}-$based superconductors, the mean-field Hamiltonian can be written as
\begin{eqnarray}
\label{h}
H&=&\sum_{\mathbf{k}}\varphi_{\mathbf{k}}^{\dag}M_{\mathbf{k}}\varphi_{\mathbf{k}},\nonumber\\
\varphi_{\mathbf{k}}^{\dag}&=&(c_{\mathbf{k}1\uparrow}^{\dag},c_{\mathbf{k}2\uparrow}^{\dag},c_{-\mathbf{k}1\downarrow},c_{-\mathbf{k}2\downarrow}),\nonumber\\
M_{\mathbf{k}}&=&\begin{pmatrix}
\varepsilon_{A\mathbf{k}}-\mu&\varepsilon_{xy\mathbf{k}}&\Delta_{1\mathbf{k}}&0\\
\varepsilon_{xy\mathbf{k}}&\varepsilon_{B\mathbf{k}}-\mu&0&\Delta_{2\mathbf{k}}\\
\Delta_{1\mathbf{k}}^{*}&0&-\varepsilon_{A\mathbf{k}}+\mu&-\varepsilon_{xy\mathbf{k}}\\
0&\Delta_{2\mathbf{k}}^{*}&-\varepsilon_{xy\mathbf{k}}&-\varepsilon_{B\mathbf{k}}+\mu
\end{pmatrix},\nonumber\\
\varepsilon_{A\mathbf{k}}&=&-2t_{1}(\cos k_{x}+\cos k_{y})-2t_{2}\cos(k_{x}-k_{y})\nonumber\\
&&-2t_{3}\cos(k_{x}+k_{y})\nonumber\\
&&-2t_{4}[\cos(2k_{x}+k_{y})+\cos(k_{x}+2k_{y})]\nonumber\\
&&-2t_{5}[\cos(2k_{x}-k_{y})+\cos(k_{x}-2k_{y})],\nonumber\\
\varepsilon_{B\mathbf{k}}&=&-2t_{1}(\cos k_{x}+\cos k_{y})-2t_{3}\cos(k_{x}-k_{y})\nonumber\\
&&-2t_{2}\cos(k_{x}+k_{y})\nonumber\\
&&-2t_{5}[\cos(2k_{x}+k_{y})+\cos(k_{x}+2k_{y})]\nonumber\\
&&-2t_{4}[\cos(2k_{x}-k_{y})+\cos(k_{x}-2k_{y})],\nonumber\\
\varepsilon_{xy\mathbf{k}}&=&-2t_{6}(\cos k_{x}-\cos k_{y})\nonumber\\
&&-2t_{7}(\cos 2k_{x}-\cos 2k_{y})\nonumber\\
&&-4t_{8}(\cos 2k_{x}\cos k_{y}-\cos k_{x}\cos 2k_{y}).
\end{eqnarray}
Here $c_{\mathbf{k}1\uparrow}^{\dag}$ creates a spin $\uparrow$ electron with momentum $\mathbf{k}$ and in orbital $1$. $\mu$ is the chemical potential, $t_{1}\cdots t_{8}$ are the hopping integrals and we consider only spin singlet intraorbital pairing up to the NN sites, thus the pairing order parameters can be expressed as
\begin{eqnarray}
\Delta_{\beta\mathbf{k}}&=&\Delta_{s\beta}+\frac{\Delta_{xs\beta}}{2}(\cos k_{x}+\cos k_{y})\nonumber\\
&&+\frac{\Delta_{d\beta}}{2}(\cos k_{x}-\cos k_{y}),
\end{eqnarray}
with $\beta=1,2$ being the orbital index and
\begin{eqnarray}
\label{delta}
\Delta_{s\beta}&=&\frac{V_{0}}{N}\sum_{\mathbf{k}}\langle c_{-\mathbf{k}\beta\downarrow}c_{\mathbf{k}\beta\uparrow}\rangle,\nonumber\\
\Delta_{xs\beta}&=&\frac{2V_{1}}{N}\sum_{\mathbf{k}}(\cos k_{x}+\cos k_{y})\langle c_{-\mathbf{k}\beta\downarrow}c_{\mathbf{k}\beta\uparrow}\rangle,\nonumber\\
\Delta_{d\beta}&=&\frac{2V_{1}}{N}\sum_{\mathbf{k}}(\cos k_{x}-\cos k_{y})\langle c_{-\mathbf{k}\beta\downarrow}c_{\mathbf{k}\beta\uparrow}\rangle,
\end{eqnarray}
are the isotropic $s-$, extended $s-$ and $d-$wave components, respectively. $N$ is the number of the lattice sites and the doping level $x$ is determined through
\begin{eqnarray}
\label{x}
x&=&\frac{1}{N}\sum_{\mathbf{k}\beta\sigma}\langle c_{\mathbf{k}\beta\sigma}^{\dag}c_{\mathbf{k}\beta\sigma}\rangle.
\end{eqnarray}

Eq. (\ref{h}) can be solved as follows: First start with a set of random $\Delta_{s\beta}$, $\Delta_{xs\beta}$, $\Delta_{d\beta}$ and $\mu$, the Hamiltonian is numerically diagonalized. Then the set of pairing order parameters and doping level are calculated by using Eqs. (\ref{delta}) and (\ref{x}) for the next iteration step and $\mu$ is adjusted according to the desired doping level ($x_{d}$). The above procedure is repeated until the absolute error of the order parameters between two consecutive steps is less than $10^{-4}$ and $|x-x_{d}|<10^{-4}$. Thus by varying the values of $V_{0}$ and $V_{1}$, the phase diagram of the pairing symmetry can be obtained. In the following, the parameters are chosen as $t_{1}\cdots t_{8}=0.167,-0.88,-0.094,-0.014,-0.069,-0.107,0.028,-0.02$, respectively. The doping level $x_{d}$ and temperature $T$ are fixed at $0.55$ and $0$, respectively. We vary $V_{0}$ ($V_{1}$) from $-0.5$ to $0$ to get the phase diagram of the pairing symmetry.

\begin{figure}
\includegraphics[width=1\linewidth]{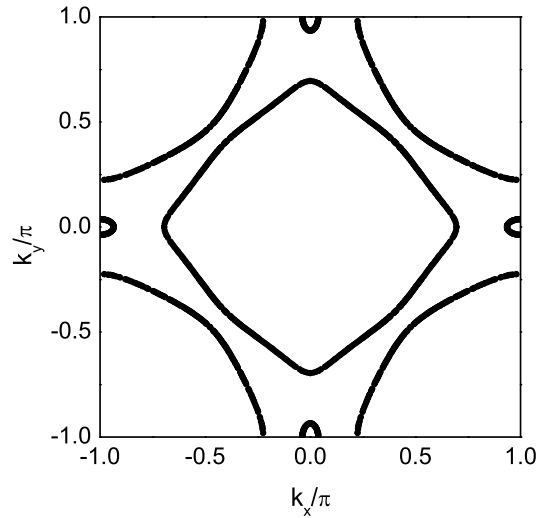}
 \caption{The Fermi surface of the two-orbital model for the doping level $x=0.55$.}
 \label{fig1}
\end{figure}

\begin{figure}
\includegraphics[width=1\linewidth]{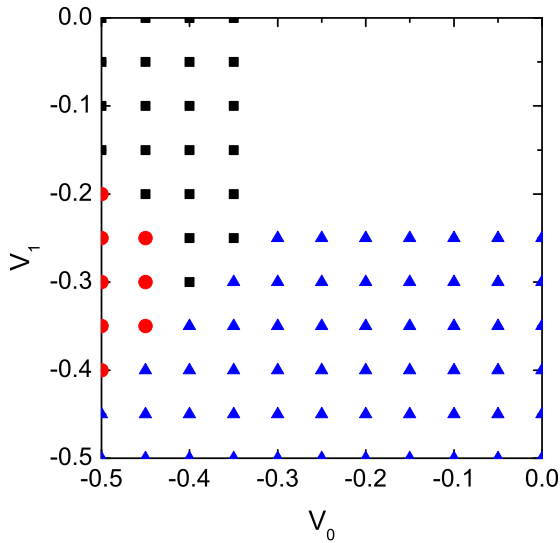}
 \caption{(color online) The phase diagram of the pairing symmetry as a function of $V_{0}$ and $V_{1}$. In the area filled with black squares, the isotropic $s-$wave pairing dominates. In the area filled with blue triangles, the $d-$wave component is dominant, whereas in the area filled with red circles, $\Delta_{s}$ with a small amount of $\Delta_{xs}$ wins over, which we denote as the anisotropic $s-$wave pairing.}
 \label{fig2}
\end{figure}

\emph{Results}.---The Fermi surface of the two-orbital model for the doping level $x=0.55$ is shown in Fig. \ref{fig1} where the small pockets around $(\pm\pi,0)$ and $(0,\pm\pi)$ emerge when $x>0.515$. The self-consistently solved order parameters satisfy
\begin{eqnarray}
\Delta_{s\beta}&=&\Delta_{s},\nonumber\\
\Delta_{xs\beta}&=&\Delta_{xs},\nonumber\\
\Delta_{d\beta}&=&\Delta_{d},
\end{eqnarray}
for $\beta=1,2$, thus the subscript $\beta$ is omitted in the following. In this case, the intraorbital pairing leads to the intraband pairing after a unitary transformation and the pairing order parameters in the band representation can still be written as
\begin{eqnarray}
\Delta_{\mathbf{k}}&=&\Delta_{s}+\frac{\Delta_{xs}}{2}(\cos k_{x}+\cos k_{y})\nonumber\\
&&+\frac{\Delta_{d}}{2}(\cos k_{x}-\cos k_{y}).
\end{eqnarray}
The phase diagram of the pairing symmetry as a function of $V_{0}$ and $V_{1}$ is shown in Fig. \ref{fig2}. We can see that the onsite and NN pairings compete with each other. When $|V_{0}|\lesssim|V_{1}|$ (the area filled with blue triangles), NN pairing wins over the onsite one and the pairing symmetry is $d-$wave. On the contrary, when $|V_{0}|>|V_{1}|$, the onsite pairing is dominant (see the area filled with black squares) and the symmetry in this case is isotropic $s-$wave. Interestingly, in a small parameter range shown as the area filled with red circles, the onsite and NN pairings coexist with each other and the symmetry is predominantly isotropic $s-$wave with a small extended $s-$wave component, thus we call it anisotropic $s-$wave pairing. Furthermore, for $|V_{0}|\lesssim0.3$ and $|V_{1}|\lesssim0.2$, no SC pairing can exist at all. The pairing phase diagram can be understood as follows: the NN attractive interaction $V_{1}$ favors $d-$wave pairing while the onsite attractive interaction $V_{0}$ leads to isotropic $s-$wave symmetry. These two pairing symmetries exclude each other, therefore there is no coexisting region of them. Only when $V_{0}$ is strong enough and $|V_{1}|$ decreases, can the extended $s-$wave pairing be induced between the NN sites, thus it must be accompanied by an isotropic $s-$wave component and cannot exist alone.

\begin{figure}
\includegraphics[width=1\linewidth]{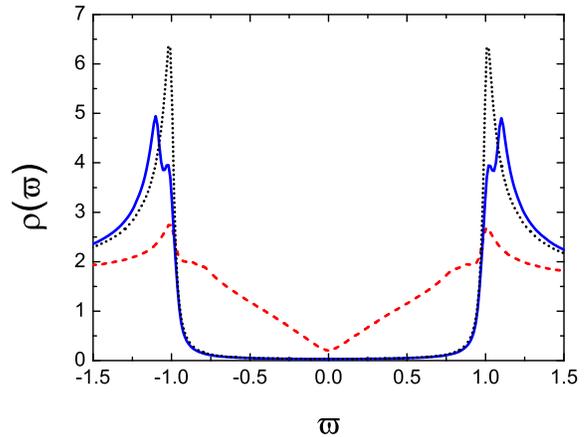}
 \caption{(color online) The DOS as a function of the reduced energy $\widetilde{\omega}$, for the isotropic $s-$ (black dot), anisotropic $s-$ (blue solid) and $d-$wave (red dash) symmetries, respectively. For the isotropic $s-$ and anisotropic $s-$wave cases, $\widetilde{\omega}=\omega/\Delta_{s}$ while for the $d-$wave case, $\widetilde{\omega}=\omega/\Delta_{d}$.}
 \label{fig3}
\end{figure}

Then we investigate the possible experimental signature of different pairing symmetries through the DOS, which can be measured by scanning tunneling microscopy (STM). The DOS is expressed as
\begin{eqnarray}
\rho(\omega)=-\frac{1}{\pi N}\sum_{\mathbf{k}\beta\sigma}Im\langle\langle c_{\mathbf{k}\beta\sigma}|c_{\mathbf{k}\beta\sigma}^{\dag}\rangle\rangle_{\omega+i0^{+}}.
\end{eqnarray}
Here $Im\langle\langle\ldots\rangle\rangle_{\omega+i0^{+}}$ stands for the imaginary part of the retarded Green's function. In Fig. \ref{fig3} we plot our calculated DOS. For the isotropic $s-$wave pairing (the black dotted line), a full gap develops in the SC state and a single pair of resonance peaks appears at $\widetilde{\omega}\approx\pm1$. For the anisotropic $s-$wave case, we choose $\Delta_{xs}\approx0.3\Delta_{s}$ and the DOS (the blue solid line) is similar to the isotropic $s-$wave case. However, around $\widetilde{\omega}\approx\pm1$, the resonance peak clearly splits into two, leading to a two-gap structure. Contrastingly, for the $d-$wave pairing, the DOS at $\widetilde{\omega}\approx0$ is finite and shows a linear dispersion, indicating the existence of nodes. Thus the different behaviors of the DOS for the three pairing symmetries can be used to distinguish them by STM experiment.

\emph{Summary}.---In summary, we have studied the possible pairing symmetries in a minimal two-orbital model for BiS$_{2}-$based superconductors, with onsite and NN intraorbital attractive interactions $V_{0}$ and $V_{1}$, respectively. By using the mean-field approximation, the phase diagram of the pairing symmetry is obtained. We found that in this model there exist three possible pairing symmetries, depending on the values of $V_{0}$ and $V_{1}$. The first one is the isotopic $s-$wave pairing [$\Delta_{\mathbf{k}}=\Delta_{s}$]. The second one is the anisotropic $s-$wave pairing [$\Delta_{\mathbf{k}}=\Delta_{s}+\frac{\Delta_{xs}}{2}(\cos k_{x}+\cos k_{y})$] and the last one is the $d-$wave pairing [$\Delta_{\mathbf{k}}=\frac{\Delta_{d}}{2}(\cos k_{x}-\cos k_{y})$]. Furthermore we propose that the DOS for these pairing symmetries exhibit different behaviors which can be used to distinguish them.

This work was supported by SRFDP (Grant No. 20123207120005) and China Postdoctoral Science Foundation (Grant No. 2012M511297).

\end{document}